# Metal-organic Frameworks in Semiconductor Devices: A revised Version

Ranjeev Kumar Parashar[1,#], Priyajit Jash[1,#], Michael Zharnikov[2,*] and Prakash Chandra Mondal[1,*]

[1]Department of Chemistry, Indian Institute of Technology, Kanpur, Uttar Pradesh-208016, India

[2]Angewandte Physikalische Chemie, Universität Heidelberg, Im Neuenheimer Feld 253, 69120 Heidelberg, Germany

E-mail: Michael.Zharnikov@urz.uni-heidelberg.de (MZ); pcmondal@iitk.ac.in (PCM)

[#]These two authors equally contributed to this work.

## Abstract

Metal-organic frameworks (MOFs) are a specific class of hybrid, crystalline, nano-porous materials made of metal-ion-based 'nodes' and organic linkers. Most of the studies on MOFs largely focused on porosity, chemical and structural diversity, gas sorption, sensing, drug delivery, catalysis, and separation applications. In contrast, much less reports paid attention to understanding and tuning the electrical properties of MOFs. Poor electrical conductivity of MOFs (~$10^{-7} - 10^{-10}$ Scm$^{-1}$), reported in earlier studies, impeded their applications in electronics, optoelectronics, and renewable energy storage. To overcome this drawback, the MOF community has adopted several intriguing strategies for electronic applications. The present review focuses on creatively designed bulk MOFs and surface-anchored MOFs (SURMOFs) with different metal nodes (from transition metals to lanthanides), ligand functionalities, and doping entities, allowing tuning and enhancement of electrical conductivity. Diverse platforms for MOFs-based electronic device fabrications, conductivity measurements, and underlying charge transport mechanisms are also addressed. Overall, the review highlights the pros and cons of MOFs-based electronics (MOFtronics), followed by an analysis of the future directions of research, including optimization of the MOF compositions, heterostructures, electrical contacts, device stacking, and further relevant options which can be of interest for MOF researchers and result in improved devices performance.

**Keywords:** metal-organic frameworks, SURMOFs, electronic devices, electrical conductivity, charge transport

# 1. Introduction

The field of metal-organic frameworks (MOFs) has experienced a tremendous evolution over recent decades, both in terms of fundamental studies and potential applications. MOFs are generally composed of metal ions or metal-ions-based clusters called 'nodes' and multitopic ligands called 'arms' or 'linkers', coupled to the 'nodes'. Bulk MOFs are commonly synthesized via a solvothermal procedure performed at high temperatures followed by slow cooling (2-5 $^{o}$C/h). During this process, 'nodes' and 'arms' coordinate with each other via a self-assembly process, leading to the formation of highly ordered, crystalline, cage-like MOFs. The journey of MOFs began in 1995 by Yaghi and team after the discovery of the first microporous framework featuring 1,3,5-benzenetricarboxylate (BTC) multidentate group as 'arms' and $Co^{2+}$ as 'nodes', held together by a coordination bond.[1,2] This thermally stable MOF was utilized for pyridine loading into its nanopores. Since then, MOFs have drawn enormous attention from the scientific community in view of their intriguing properties and diverse applications. Large pore size and design diversity make these systems promising candidates for storing small molecules, gas absorption media, sensing platforms, catalytic materials, photo-responsive systems, and biomedical applications.[3–6] At the same time, their poor electrical conductivity, arising generally from weak *d-p* orbitals interaction between transition metal ions and organic linkers, makes them unfavorable for electronic and energy storage applications. A variety of diverse strategies has been implemented to enhance the electrical conductivity of MOFs, including the improvement of node-ligand coupling, the use of extended conjugated ligands, control over π-π stacking of the ligands, introduction of mixed valency compounds, the use of redox-active metals, ions, and ligands, and the loading of guest compounds, such as different molecules, conductive polymers, and polar solvents into the void space of MOFs. As the result, the conductivity could be tuned and enhanced significantly compared to the initial, poorly conductive MOFs. Besides the above factors affecting the material properties of a particular MOF, there are several other factors which should be considered and optimized upon the integration of MOFs into electronic devices, such as bias-induced geometry around the metal centres, temperature, thickness and anisotropy of the MOF films, nature of the electrodes, electronic coupling between the electrodes and MOF, and exact device configuration. Also, understanding the relevant charge transport mechanisms is still an issue, frequently assisted by temperature-dependent current-voltage (*I-V*) data that give the value of the activation energy ($E_a$) required for the charge transport inside the MOFs. There is a variety of mechanistic charge transport models, such as coherent tunnelling, thermally activated hopping, band transport, and redox-hopping (see below). The situation in MOF films, which are usually thicker than 5 nm, is different from that in the thinner molecular films and is also different from that in organic semiconductors.[7,8] Also, recent experimental studies showed that selected lanthanide-based MOFs and redox-active ligand-containing MOFs represent better conductors than the conventional frameworks.[9–11] In particular, Dincă and co-workers demonstrated that the lanthanides-MOFs with suitable ligands feature extensive charge delocalization, revealing π-π stacking based conductivity as high as 1,000 Scm$^{-1}$, close to metallic behaviour.[9] We discuss these systems in the present review, along with some other recently emerged systems showing a promising potential for electronic applications, such as metal-catecholate based MOFs, Ni-HITP MOFs (HITP = 2,3,6,7,10,11-hexaaminotriphenylene), and Ga-based MOFs. The emphasis is put on the thin film MOFs and such relevant issues as the anisotropy of electrical conductance, tuning of conductance by external stimuli, its modification by doping, and the design of MOF-based devices for specific electronic applications. Except for these major issues, we shortly discuss such general points as the current models of electrical conductivity in MOFs, experimental approaches to the fabrication of thin film MOFs, and standard experimental techniques to measure the electrical conductance.

More detailed information on these points and some other relevant questions in the general context of electrically conductive MOFs can be found in the dedicated reviews and perspectives published recently in various journals.[12–16] One can probably especially emphasize the recent review by Dincă and co-authors providing comprehensive information on electrically conductive MOFs,[12] a review by R. Freund *et al.* describing the current state of designing and tuning MOF and COF (coordination polymer framework) properties and the respective electronic applications,[14] a review by Dincă and co-authors highlighting ion and electron transport in MOFs,[15] and a perspective by J. Park and co-authors on band gap and electronic structure tuning in 2D MOFs.[16] A particular interesting example of such a MOF, featuring both high conductivity and a record-high surface area, was recently reported by Park and co-workers.[17] A representative list of the earlier reviews and perspectives can be found in Ref.[12].

## 2. Design and fabrication of thin-film MOFs

A schematic overview of a roadmap in context of conductive MOFs and the relevant fabrication techniques for thin MOF films are presented in **Fig. 1**. Progress in the synthesis of MOF components, design and engineering of MOFs, and solution-based MOF chemistry has propelled a promising research domain called 'SURMOFs', led, among others, by the team of Christof Wöll.[18–20] SURMOFs are thin, crystalline MOF films grown on the solid substrates.[21–23] Recent progress in general strategies for preparation of MOFs with different dimensions, including SURMOFs in particular, is well described in the recent reviews by Pang and co-authors and Wöll and co-authors.[24,25] The growth of SURMOFs is frequently performed by layer-by-layer (LbL) method, starting from a template layer formed by pre-functionalization of the substrate (**Fig. 2a**).[25–29] The pre-functionalization step involves the formation of a suitable self-assembled monolayer (SAM), which can serve as a chemical template for the subsequent SURMOF growth.[30] The successive layers are usually applied by liquid-phase epitaxy (LPE).[31,32] Other methods for SURMOF growth include spin coating, drop-casting, chemical vapor deposition, atomic layer deposition, substrate-seeded heteroepitaxy, and electrochemical deposition in which redox-active ligands and metal ions precursors are introduced into an electrochemical cell (**Fig. 2b-d**).[31,32] Among these techniques, LPE was found to be particularly favorable as it not only enables the formation of homo SURMOFs but heterostructures as well and provides well-defined electrical contacts to electrodes, crucially important for the device performance.[28,33] During the LPE process, performed usually in the LbL fashion, the building blocks of SURMOFs are assembled successively and individually compared to the ''one-pot'' reaction scheme applied for the solvothermal synthesis of bulk MOFs and to the spin-coating and drop-casting strategies applied for the preparation of SURMOFs (**Fig. 2**). A representative example of a SURMOF fabricated by the LPE process is HKUST-1 - a well-known framework comprised of copper ions and trimesic acid (TMA), prepared as a highly oriented crystalline film with variable thickness on a carboxylic acid terminated SAM template on the Au substrate.[34–36] Note, however, that SAMs are generally not only useful as SURMOF templates but also represent a versatile platform for a broad variety of experimental and theoretical studies as well as for nanotechnological applications.[37–40] Several research groups have reported the fabrication of SAM-based molecular electronic and spintronic devices incorporating molecules and MOFs as circuit elements.[39,41–45]

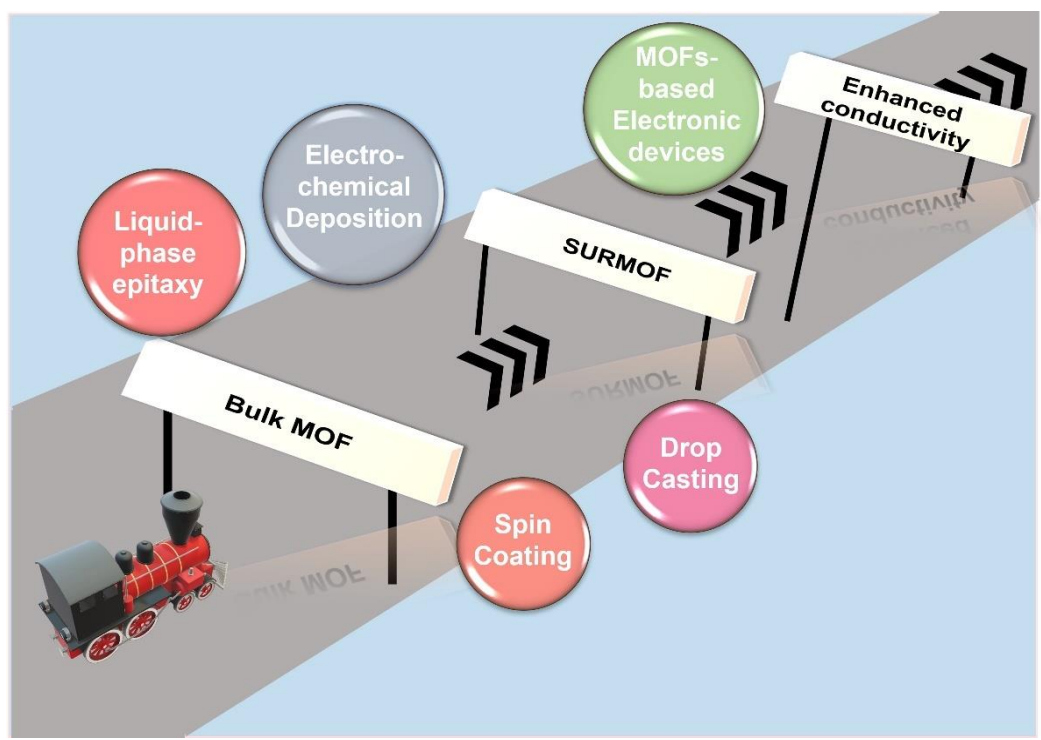

**Figure 1.** Schematic representation of the roadmap on the way to MOF-based electronic devices as well as diverse routes for MOF assembly on solid substrates used in the given context.

The electrochemical deposition technique is another simple and robust strategy for the SURMOF growth. It can be categorized into three methods: (i) anodic deposition, (ii) cathodic deposition, and (iii) electrophoretic deposition. The key advantage of the electrochemical approach is that it allows fast growth of homo and heterostructures with variable compositions and thicknesses, suitable for electronic devices.[46–49]

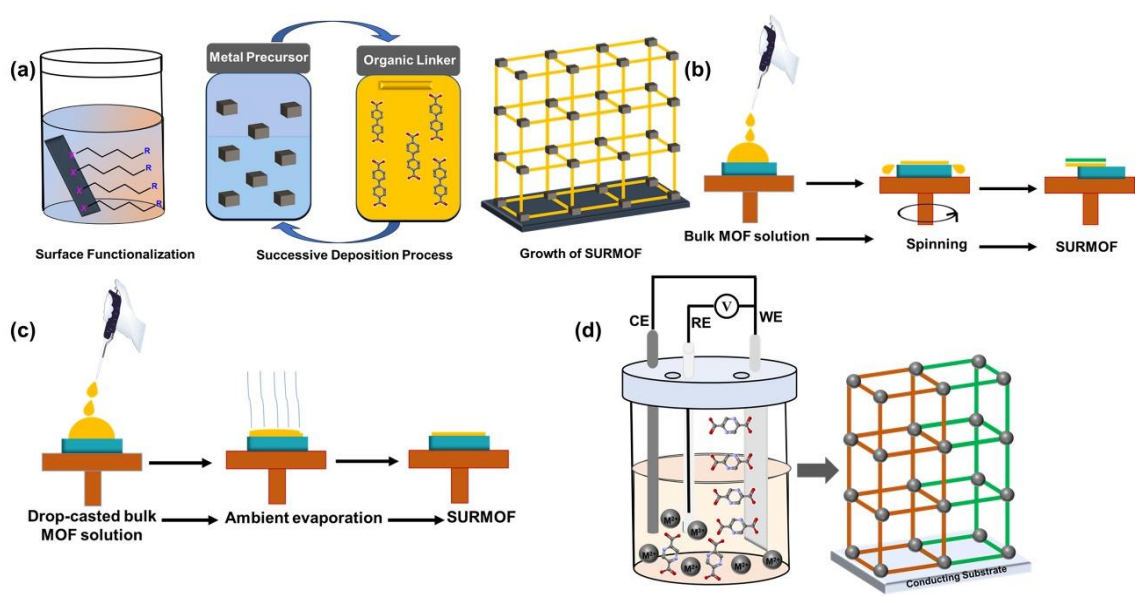

**Figure 2.** Schematic illustration of the most popular techniques for fabricating SURMOFs on pre-functionalized substrates. The techniques include (a) LPE-LbL assembly of metal ions and organic linkers on a SAM-engineered substrate, (b) spin-coating, (c) drop-casting, and (d) electrochemical deposition. Vapor deposition is not shown.

## 3. Breaking the bottleneck of electrical conductivity in MOFs

Despite successful use of MOFs in various electronic applications and devices, it remains a challenging task to understand an underlying charge conduction mechanism. The majority of MOFs feature a wide band gap, and the orbitals of the metal ion clusters, and redox-inert linkers do not overlap sufficiently to allow for effective charge transport in the framework. As a result, most of the MOFs are electrical insulators. The first electrically conductive MOF, $Cu^I[Cu^{III}(pdt)_2]$ (pdt = pyrazine-2,3-dithiolate], was reported in 2009 by Takaishi and co-workers.[50] This MOF utilized mixed metal ions with low and high oxidation states ($Cu^I$ and $Cu^{III}$) acting as electron donors and acceptors, respectively, and enhancing electrical conductivity. Even though the conductivity value, $\sigma = 6 \times 10^{-4}$ Scm$^{-1}$, was rather moderate, that work laid a foundation stone for developing further conductive MOFs and their integration with other elements for investigating charge transport phenomena and applications. Since then, significant efforts have been made towards MOF-based multifunctional electronic devices, which provide us with a new research domain popularly known as ''***MOFtronics***'' (MOF-based electronics). The evolution of conducting MOF for the period of 2009-2023 is sketched in **Fig. 3**. It reflects the progress in the field due to the development of different "tools" to tune and enhance the conductivity, including the efficient coupling of nodes to linkers, doping with redox-active guest molecules, inclusion of polar solvents, use of different metal ions (mixed valency), and the introduction of lanthanides as MOF "nodes".

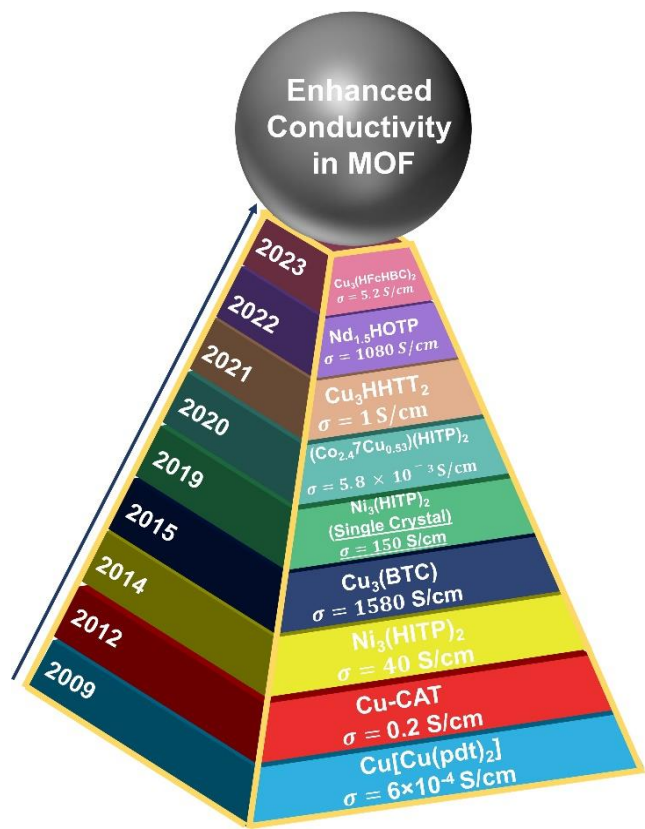

**Figure 3.** Evolution of electrically conductive MOFs for the period from 2009 to 2023. Note that this is not only the absolute value of the conductivity which is of importance but also introduction of new MOF systems, opening new perspectives for research and applications. The factors that facilitate electrical conductivity include doping, mix-metal ions, conductive linkers, improvement of π-π stacking, conformational changes, mixed valency, etc.

## 4. Measurements and models of electrical conductivity in MOFs

It is generally assumed that the charge transport in conductive MOFs occurs in either "through-bond" or 'through-space' fashion, relying on either band or hopping transport mode (**Fig. 4**).[51] There are also several other models dealing with specific cases.[52] Note that "through-bond" charge transport is related to the flow of charges over metal-ligands coordination bonds; it propagates through the orbitals of ligands and those of metal ions (**Fig. 4a**). On the other hand, 'through space' conduction channels originate from non-covalent interactions between ligands. For instance, MOFs composed of extended π-conjugated linkers, like triphenylene, trinaphthylene, hexaazatrinaphthylene, phthalocyanine, and naphthalocyanine, feature strong π−π interactions between the ligands, facilitating charge transport via 'through space' pathways.[53] In contrast, 'through bond' charge transport relies on a strong covalent bonding between metal centres and organic linkers, resulting in a small band gap and high charge mobility. In particular, this mechanism is favoured for nitrogen and sulphur-based linkers featuring well-matched energy levels and spatial overlap with the metal d-orbitals.[54] As to the hopping transport, it is a thermally activated process with charge carriers localized at specific sites and 'jumping' from these sites to other sites (**Fig. 4b**).[55] The conductivity depends then strongly on temperature, as is typical for activated processes. A detailed analysis of the aforementioned charge transport mechanisms can be found elsewhere.[13,15,54,56,57]

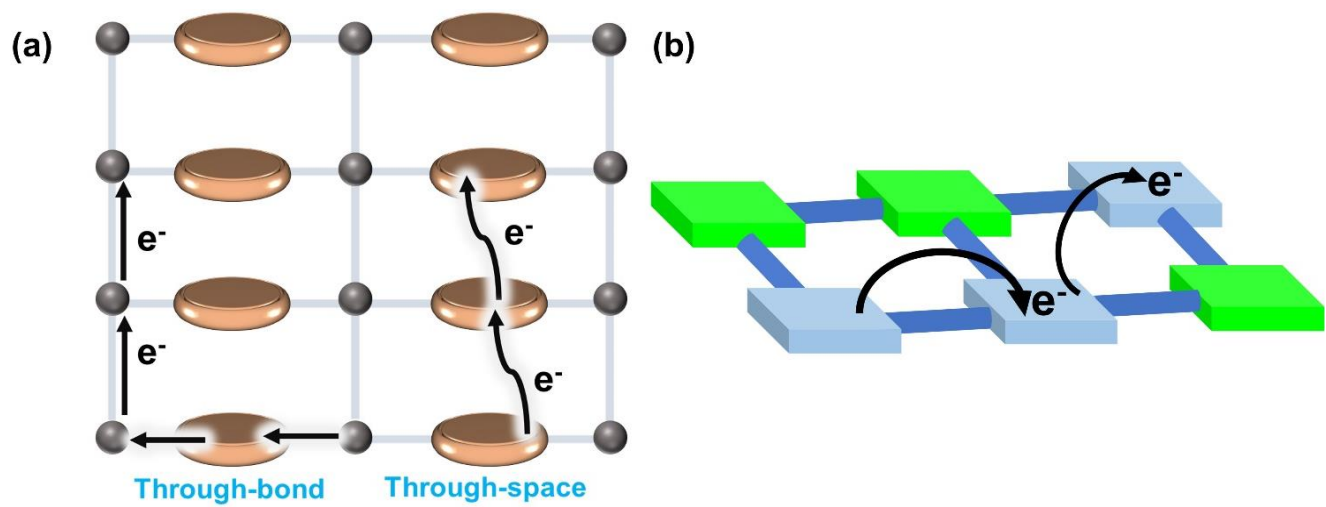

**Figure 4.** (a) Schematic illustration of the 'through-bond' and 'through-space' charge conduction mechanisms in electrically conductive MOFs; (b) Schematic illustration of the hopping transport.

### 4.1 Measurement of electrical conductivity and its anisotropy

Charge transport properties of MOFs are best represented by their electrical conductivity, σ, a parameter that reflects the ease with which charges can propagate through MOFs or other materials, can be calculated according to the equation,

$$\sigma = \frac{I}{V} \times \frac{L}{A} \qquad (i)$$

where $I$, $V$, $L$ and $A$ denote current, applied bias, the distance between the two electrodes (thickness of a MOF film), and the cross-sectional area of the junction, respectively.

Conductivity of MOFs can be measured for both single crystals and polycrystalline materials in the form of either thin films or pellets. In polycrystalline samples, grain boundary resistance and distribution of different crystallographic orientations contribute to the conductivity in addition to their inherent conductivity which is the property of fundamental interest. Nevertheless, most of the charge transport measurements so far were performed on polycrystalline thin film MOFs or pellets, which are the most frequently used MOF samples. At the same time, measurements on single-crystal MOFs have their own importance since they provide the most characteristic material values, serving as references for polycrystalline samples and as a benchmark for applications.

The charge transport measurements are usually performed by either two-probe or four-probe methods.[13] As electrical contacts, metal wires (Ag, Au, Cu, or Al) or conductive adhesives (carbon paste, gold, or silver paint) are usually employed. For instance, Dincă and co-workers measured the conductivity of the $Cd_2$(TTFTB) MOFs by a two-probe method, using conducting carbon paste and gold wires to make the electrical contacts.[58] This method involves applying a known voltage or current to the sample and measuring the resistance (**Fig. 5a**). Its major drawback is the contribution of the contacts, wire resistance, and voltage drop which can be avoided by four-probe method that exclusively measures the intrinsic conductivity of the sample (**Fig. 5b-c**). An extension of the four-probe method is the van der Pauw method, which is used when the sample is very thin or irregular in form (**Fig. 5d**).[56]

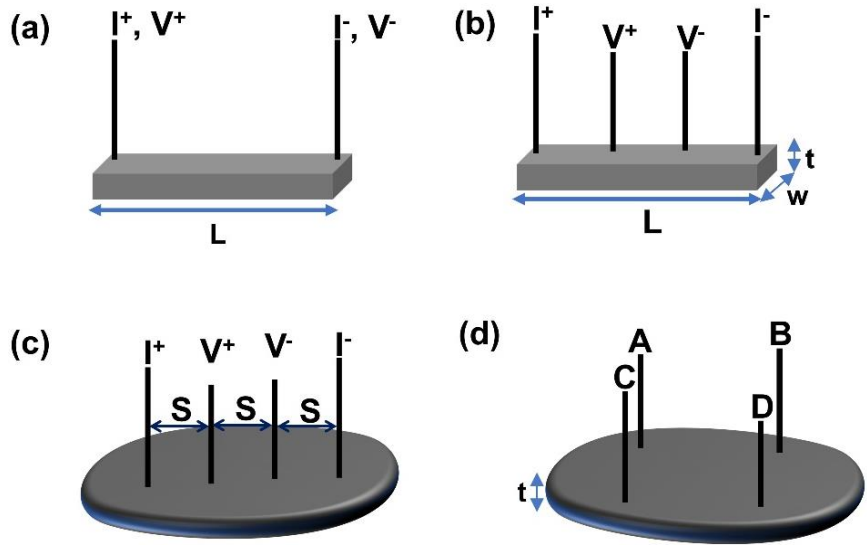

**Figure 5.** Schematic diagram of the methods used to measure electrical conductivity of MOFs. (a) Two-probe method, (b-c) four-probe method, (d) van der Pauw method. *I*, *V*, *L*, *W*, *t*, and *S*, indicate current, voltage, length, width, and thickness of the sample, and the distance between the contacts, respectively.

In the context of single-crystal MOFs, we generally consider in-plane and out-of-plane conductivities. In-plane conductivity typically refers to the coordination bonds within the same plane of the MOF structure. In contrast, out-of-plane conductivity entails charge carriers moving perpendicular to the primary planes, often involving charge transport via hopping, or tunneling mechanism. The major challenges to acquire these parameters are to prepare single-crystal MOF samples with appropriate orientations and to contact them properly relying on the suitable crystal face. Another critical issue for the electrical conductivity measurements on single-crystal MOFs is the reproducibility of the results, affected both by the distribution of the parameters of the crystals (size, shape, purity, and morphology) and by the quality of the contacts. Hence, smart tactics are required to meet the above challenges. Toward this endeavor, Sun *et al.* probed the electrical conductivity of a single crystal of $Cd_2(TTFTB)$ MOF with a hexagonal shape for both in-plane ($\sigma\|c$) and out-of-plane ($\sigma\perp c$) directions, relying on the specific faces of this crystal.[58] The authors attached gold wires with tungsten probes to preclude the crystal from being hampered. The obtained conductivity values for $\sigma\|c$ and $\sigma\perp c$ orientations were estimated at 9.82 ($\pm$2.09) × $10^{-5}$ S/cm and 2.36 ($\pm$0.72) × $10^{-7}$ S/cm, respectively.[58] Along the same line, Dou *et al.* investigated in detail the electrical conductivity of an array of 2D anisotropic $M_mHHTT_n$ MOFs, with either M = $Cu^{2+}$ or $Ni^{2+}$, m = 3, and n = 2 or M = $Mg^{2+}$, $Co^{2+}$ or $Ni^{2+}$, m = 6, and n = 3.[59] The in-plane and out-of-plane electrical conductivity measurements were performed on hexagonal single-crystal plates and single-crystal rods, respectively. As expected, the porous, eclipsed phase $M_3HHTT_2$ (M = $Cu^{2+}$ or $Ni^{2+}$) exhibited higher in-plane than out-of-plane conductivity. In contrast, the dense staggered phase $Co_6HHTT_3$ showed higher out-of-plane conductivity, which was explained by the presence of planar $Co_3$ clusters in this framework. It was assumed that these clusters preferentially facilitate charge transport perpendicular to the 2D sheets while concurrently preventing the transport within the sheet via narrower band dispersion or greater phonon scattering. Furthermore, for all MOFs studied, the out-of-plane conductivity was inversely proportional to the interlayer π–π distances, except for $Mg_6HHTT_3$, which was explained by the absence of d orbital in its metal ions. Recently reported values of the orientation-specific electrical conductivities of single crystal MOFs are summarized in Table 1. The common trend is that in-plane conductivity is higher than out-of-plane and pellet values, which is understandable in view of the MOF structure.

A particularly high value of in-plain conductivity in the single-crystal MOFs, 10.96 S cm$^{-1}$, was measured for the single crystals of the Cu-based frameworks, {[$Cu_2$(6-Hmna)(6-mn)]·$NH_4$}$_n$ (6-Hmna = 6-mercaptonicotinic acid, 6-mn = 6-mercaptonicotinate) featuring 2D copper–sulfur planes.[60] The key approach was to expand the (–Cu–S–)$_n$ chains to (–Cu–S–)$_n$ planes, which permits efficient charge transport. A particular high value of the out-of-plain conductivity, such as 150 S cm$^{-1}$, was measured in the single crystals of $Ni_3$(HITP)$_2$ MOFs [3] – the system which will be discussed in detail below.

**Table 1.** Orientation-specific electrical conductivities of the single-crystal MOFs. Measurement conditions, techniques, and conductivity values for the respective polycrystalline samples (pellets) are provided.

| Examples of MOFs | Measurement technique | Measurement conditions | Conductivity (S cm$^{-1}$) | Orientation | Ref. |
|---|---|---|---|---|---|
| $Cd_2$(TTFTB) | Single crystal (2-probe) | $N_2$-atmosphere (Glove box) | $10^{-4}$ | ab-axis | Sun et al. [58] |
| $Cd_2$(TTFTB) | Single crystal (2-probe) | $N_2$-atmosphere (Glove box) | $10^{-7}$ | c-axis | Sun et al. [58] |
| $Cd_2$(TTFTB) | Pellet (2-probe) | $N_2$-atmosphere (Glove box) | $10^{-6}$ | Random | Sun et al. [58] |
| $Ni_9$(HHTP)$_4$ | Single crystal (4-probe) | Vacuum | 2 | ab-axis | Ha et al. [61] |
| $Ni_9$(HHTP)$_4$ | Single crystal (2-probe) | Vacuum | $1\times10^{-4}$ | c-axis | Ha et al. [61] |
| $Ni_9$(HHTP)$_4$ | Pellet | Vacuum | $3.6\times10^{-3}$ | Random | Ha et al. [61] |

| | (4-probe) | | | | |
|---|---|---|---|---|---|
| Cu$_3$HHTT$_2$ | Single crystal (vdP) | Vacuum | ~ 1 | ab-axis | Dou et al.[59] |
| Cu$_3$HHTT$_2$ | Single crystal (4-probe) | Vacuum | ~ 0.1 | c-axis | Dou et al.[59] |
| Cu$_3$(HHTP)$_2$ | Single crystal (4-probe) | Vacuum | 1.5 | ab-axis | Day et al.[3] |
| Cu$_3$(HHTP)$_2$ | Single crystal (2-probe) | Vacuum | 0.5 | c-axis | Day et al. [3] |
| Cu$_3$(HHTP)$_2$ | Pellet (2-probe) | Vacuum | 0.1 | Random | Day et al. [3] |
| Cu-Hmna | Single crystal (4-probe) | Ambient | 10.96 | ab-axis | Pathak et al.[57] |
| Ni$_3$(HITP)$_2$ | Single crystal (4-probe) | Vacuum | 150 | c-axis | Day et al. [3] |
| Ni$_3$(HITP)$_2$ | Pellet (4-probe) | Ambient | 55.4 | Random | Chen et al. [62] |

## 5. Bias-induced structural modulation yielding different conductivity in MOFs

External stimuli, such as electric or magnetic fields, light, temperature, and pressure, can significantly influence the charge transport properties of MOFs. For example, McGrail and co-workers prepared conductive MOFs composed of the TCNQ (7,7,8,8-tetracyanoqunidodimethane) linkers and Cu$^{2+}$ nodes which could reversibly change the structural configuration in response to applied potential.[63] These structural changes induce either a conducting state (phase I; σ = 4.8× 10$^{-3}$ S cm$^{-1}$) or an insulating state (phase II; σ = 5.8× 10$^{-7}$ S cm$^{-1}$). These MOFs (phase II) were incorporated in electronic devices by placing them between the Al foil and Cu contacts (**Fig. 6a-b**). The corresponding current-voltage (*I-V*) curve indicates negligible current on biasing the junction from 0 to +1.3 V. In contrast, at a higher bias of +6 V, the devices produce significant current with a little effect of the scan rates. It was found that the insulating phase II of Cu(TCNQ) features N−Cu−N angles of 103° and 114.7° which corresponds to a nearly tetrahedron geometry around the Cu centres. In this structure, TCNQ linkers are parallel to each other but have a sizeable interplanar distance of 6.8 Å, which is too large for efficient charge transport across the framework (**Fig. 6c**, right). In contrast, the conducting phase I exhibits N−Cu−N angles of 92º and 142º corresponding to an extremely distorted tetrahedron structure, where the neighbouring TCNQ molecules are arranged perpendicular to each other with an interplanar stacking distance of 3.24 Å, which is shorter than the van der Waals distance of carbon atoms of 3.4 Å (**Fig. 6c**, left). As the result, TCNQ molecules in phase I create a columnar stack with efficient π-π interaction, while phase II cannot form π-π stacking due to the larger distance between the neighbouring TCNQ linkers. These results agree qualitatively with the earlier study by another group reporting a field-induced conductance switching from a high (~100 kΩ) to low (~100 Ω) resistance state in the same system.[64] According to this study, the structure of Cu(TCNQ) MOFs exhibits a phase transformation occurring at +6 V, with an intermediate structural change at +3 V and (**Fig. 6c**). This transformation causes a change of the largest and lowest angles of CN- groups around the Cu centre, resulting in different TCNQ stacking, and porosity. These changes facilitate interaction between the d orbitals of the metal centres and the π orbitals of the cyano-bridges, resulting in enhanced electrical conductivity.

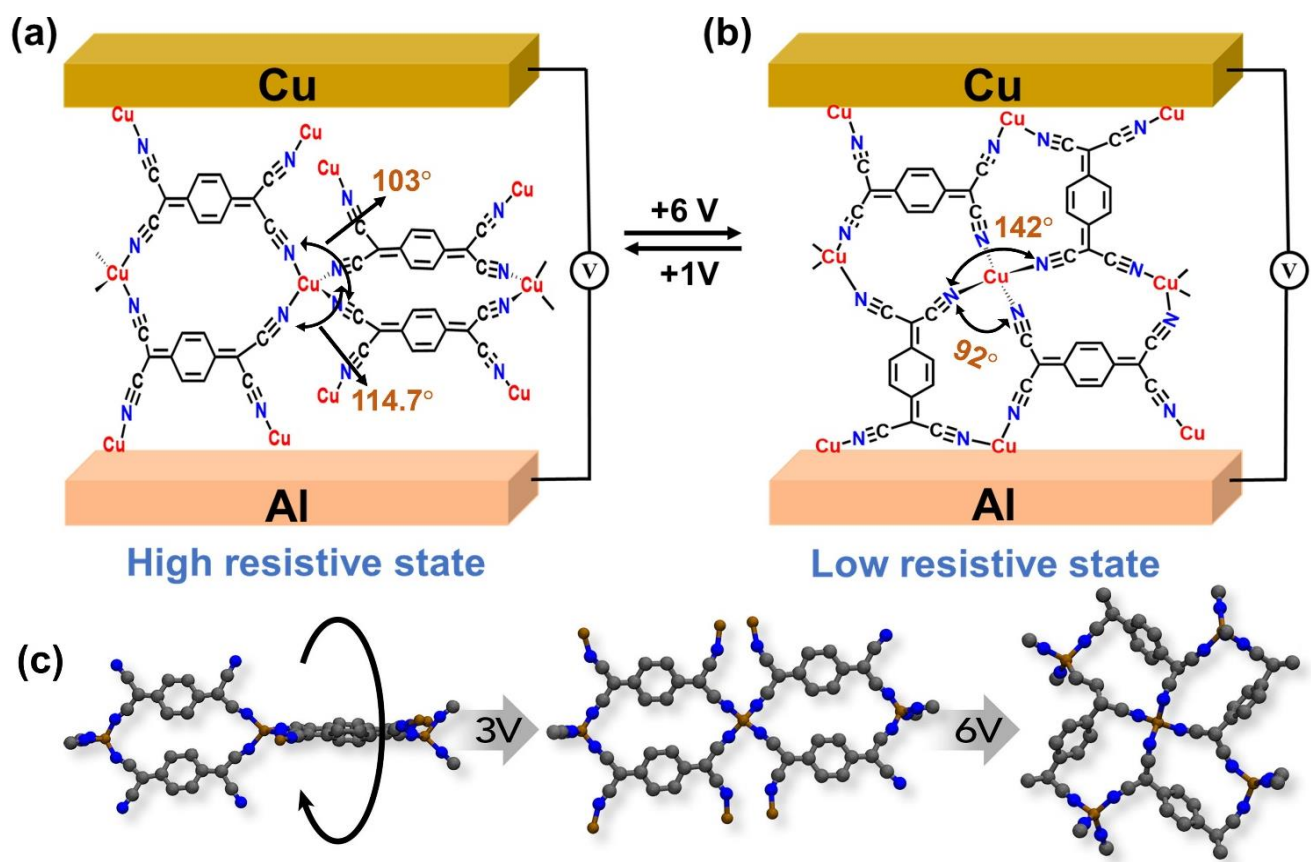

**Figure 6.** Bias-induced reversible conductivity switching of MOFs-based devices, Al/Cu(TCNQ)/Cu, between the high (a) and low (b) resistive state (phases I and II, respectively). (c) schematic illustration of the respective structural changes in the MOFs. Color code: gray: carbon, blue: nitrogen, and brown: copper. Phase I features a largely distorted tetrahedron around the Cu centre (left), while phase II represents a nearly undistorted tetrahedron (right). Reprinted with permission from Ref.[63]. Copyright 2015, Springer Nature.

## 6. Photoconductive MOFs

Light represents a versatile and non-destructive stimulus to modulate the electronic properties of photo-responsive MOFs, relying in particular on photosensitive linkers, facilitating charge transport.[65] Accordingly, light-driven charge transport or photo-conduction is an important domain of MOFtronics, with such potential applications as photoswitches, photosensors, photoelectrodes, photocatalysis, and memory devices.[66–71] A representative building block in this context is spiropyran exhibiting distinct structural and conformational changes triggered by light.[72–75] Specifically, a neutral spyropyran (SP) can transform into charge-separated zwitterionic form, commonly known as merocyanine (MC) and featuring specific electrical properties. Accordingly, spyropyran was assumed to act as a photolinker in $Zn^{2+}$-based MOFs with reversible photoconduction. To this end, Shustova and co-workers prepared photochromic MOFs featuring spiropyran as the linkers (**Fig. 7a**).[76] These MOFs changed their color from green to red upon excitation with UV light (365 nm), manifesting the transformation of the SP into MC form, featuring an extended π-conjugation (**Fig. 7b**). Accordingly, a significant increase in electrical conductivity was observed, accompanied by a correlation between the conductance and light absorption behaviour upon the repetitive photoisomerization cycles (**Fig. 7c**). However, the exact change in the electrical conductivity was not possible to deduce due to fast cycloreversion of the MC to SP form. The respective rate constants for the UV-induced transformation ($k_{UV}$) and cycloreversion ($k_{vis}$) were found to be $7.6 \times 10^{-2}$ s$^{-1}$ and $2.2 \times 10^{-2}$ s$^{-1}$, respectively. The observed photoconductivity behaviour was additionally visualized by integrating the SP-containing MOF pellet as a serial resistor into a single-transistor amplifier circuit featuring a LED. As expected, the LED shone upon irradiation of the MOF with UV light (ON state) while it got dark (OFF state) once the light was switched off (**Fig. 7d, e**). The same authors have also investigated the photoconductivity of diarylethene-containing MOFs.[76] Diarylethene is known to undergo a light-driven transformation from the 'open' to 'closed' structure, which can be in particular traced by x-ray photoelectron spectroscopy (XPS), with the characteristic S 2p$_{3/2}$ signal at 164.1 eV and 163.4 eV for the 'open', and 'closed' forms, respectively. Following such a transformation, the conductivity of the pristine, diarylethene-containing MOFs, $9.5 (\pm 2.1) \times 10^{-7}$ Scm$^{-1}$, increased to $2.9 (\pm 0.67) \times 10^{-6}$ Scm$^{-1}$ upon UV irradiation.

Another useful building block in the context of optoelectronic applications is porphyrin which is an excellent electron donor with a delocalized π-system and absorption in the visible region.[20] Additionally, the Fermi level of porphyrin can be easily modulated by its coordination with different metals; thereby, S-scheme heterojunctions can be prepared by employing it. For instance, most recently, Zhu and co-workers discovered that a heterojunction consisting of an ultrathin film of positively charged CuTCPP(Cu) and negatively charged CuTCPP(Fe), generates a higher photocurrent than the respective individual MOFs.[77] The authors concluded that the high photocurrent in the heterojunction is due to the formation of S-scheme heterojunction, leading to the spontaneous movement of electrons from CuTCPP(Cu) to CuTCPP(Fe) due to the differences in their Fermi energy levels. In addition, the band bending at the CuTCPP(Fe)/CuTCPP(Cu) interface creates a huge driving force for efficient charge carrier separation. Consequently, recombination of photogenerated carriers is suppressed. This smart heterojunction design paves the way for the fabrication of highly efficient, MOF-based optoelectronic devices for practical applications such as photosensors, photoswitches, photoelectrodes, etc.

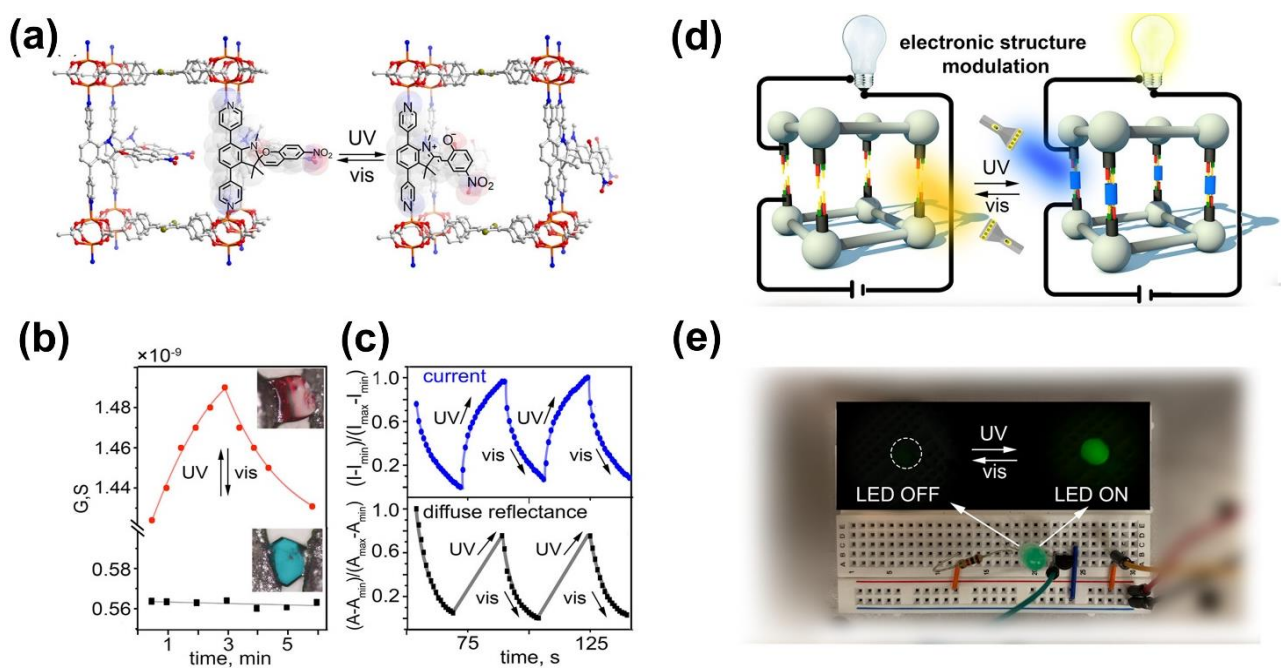

**Figure 7.** (a) Photoisomerization of spiropyran (neutral form) to merocyanine (charge separated form) in a MOF; orange, red, blue, brown, and gray spheres represent zinc, oxygen, nitrogen, bromine, and carbon atoms, respectively; (b) light-driven conductance switching in a photoresponsive (red curve) and reference (black curve) MOFs; (c) reversible photoconductance cycles of current (top panel) and corresponding absorption signal (bottom panel); (d) schematic illustration of photo-driven electronic modulation in MOFs; and (e) demonstration of enhancement in the electrical conductivity of the MOFs through LED lighting. Reprinted with permission from Ref.[76] Copyright 2019, American Chemical Society.

A promising approach to rational design of functional photoswitchable MOFs was demonstrated recently by M. Mostaghimi et al.[78] Using multiscale modelling, automated workflow protocols, and DFT calculations, they analyzed and optimized a family of model MOFs incorporating spiropyran photoswitches at controlled positions. It was shown that lattice distances and vibrational flexibility strongly modulate the possible conduction photoswitching. The electronic coupling between the HOMO and LUMO orbitals of the linkers in the tested MOFs was found to increase on average more than 23 and 65 times upon spiropyran-to-merocyanine isomerization, corresponding to the on/off ratios of 530 and 4200, respectively. Additional decoration of the photoswitches with electron-donating or -withdrawing groups opens a way to controlled modification between electron and hole conduction in a MOFs.

## 7. Tuning conductivity of MOFs using doping entities

Doping entities can effectively tune and significantly improve the electrical properties of MOFs. For instance, Farha and co-workers reported doping of the NU-1000 MOF, formed by pyrene-based linkers and hexa-zirconium nodes, with nickel (IV) bis(dicarbollide) (NiCB).[79] The electrical conductivity was measured by the placement of NU-1000, NiCB, and NU-1000@NiCB into two-terminal junction featuring Pt interdigitated electrodes (IDEs). In the resulting *I-V* curves, zero slope was obtained for both NU-1000 and NiCB, whereas a non-zero slope was measured for NU-1000@NiCB, corresponding to an electrical conductivity of $2.7\times10^{-7}$ S/cm. The remarkable enhancement of the conductivity was attributed to the donor-acceptor charge transfer between the pyrene-based linkers and NiCB guests. Similarly, conducting polymer fibres embedded in MOFs lead to a significant increase in electrical conductivity,[80,81] which is generally attributed to charge transfer interactions between polymer chains surrounded by π-donor ligands as well as to high degrees of order and orientation in the polymers.[54] For instance, the UiO-66 MOF is an insulator with a conductivity of ~$10^{-8}$ S cm$^{-1}$ (**Fig. 8a**).[82] However, a substantial rise in the electrical conductivity was observed when this framework was loaded with polypyrrole (PPy) as well as poly 3,4-ethylenedioxythiophine (PEDOT), while preserving porosity comparable to the undoped MOF, which was evident from the electrical conductivity values and surface area measurements (**Fig. 8b**). As a result, these doped MOFs emerge as compelling contenders for energy storage devices, such as supercapacitors, batteries, etc.

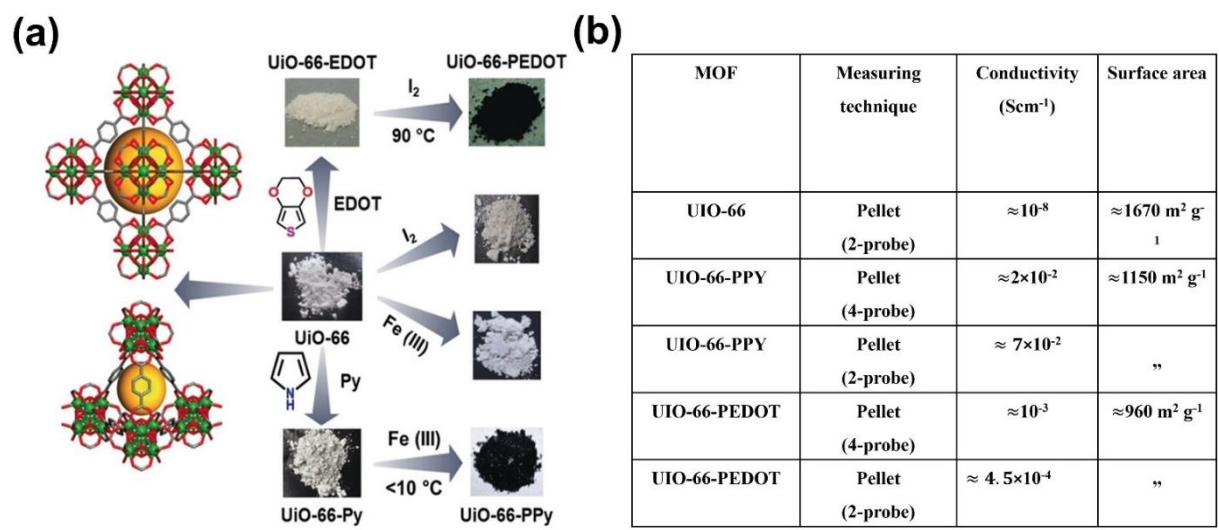

**Figure 8.** (a) Schematic illustration of the fabrication of the UiO-66-PPY and UiO-66-PEDOT MOFs, where UiO-66-PPY was synthesized by the loading with PPy followed by the oxidation with FeCl$_3$ and UiO-66-PEDOT was synthesized by the loading with PEDOT followed by the oxidation with I$_2$; color code: green: zirconium, red: oxygen, gray: carbon; hydrogen atoms are omitted for clarity. (b) The electrical conductivity and surface area values for the doped and undoped (UIO-66) MOFs. Reprinted with permission from Ref.[82] Copyright 2020, Wiley-VCH.

Alternative to doping with conducting polymer fibres, *in situ* polymerization within the frameworks can be performed. To this end, Ballav and co-workers reported the inclusion of monomer pyrrole (Py) into the nanochannels of [Cd(NDC)$_{0.5}$(PCA)] MOFs (NDC = 2,6-napthalenedicarboxylic acid and PCA = 4-pyridinecarboxylic acid].[83] The subsequent treatment of these MOFs with iodine (I$_2$) to form polypyrrole (PPy) within the framework resulted in color changes considered as an indication of the polymerization (**Fig. 9 a-b**).[83] **Fig. 9c** shows field-emission scanning electron microscopy (FE-SEM) images which prove a similar morphology of the [Cd(NDC)$_{0.5}$(PCA)] and [Cd(NDC)$_{0.5}$(PCA)]PPy MOFs. Energy dispersive X-ray spectra suggest the presence of nonstoichiometric and insignificant amounts of $I_2$, corroborating its role as an initiator for oxidative polymerization of pyrrole.[84] As the result of this polymerization, the electrical conductivity of the pristine [Cd(NDC)0.5(PCA)] MOFs (~$10^{-12}$ Scm$^{-1}$) increased by nine orders of the magnitude, becoming ~$10^{-3}$ Scm$^{-1}$ (**Fig. 9d**). An additional evidence for the positive effect of conductive polymers was provided by Dincă and co-workers, who have reported that insulating MOFs loaded with conductive polymers turn into highly conductive frameworks.[12]

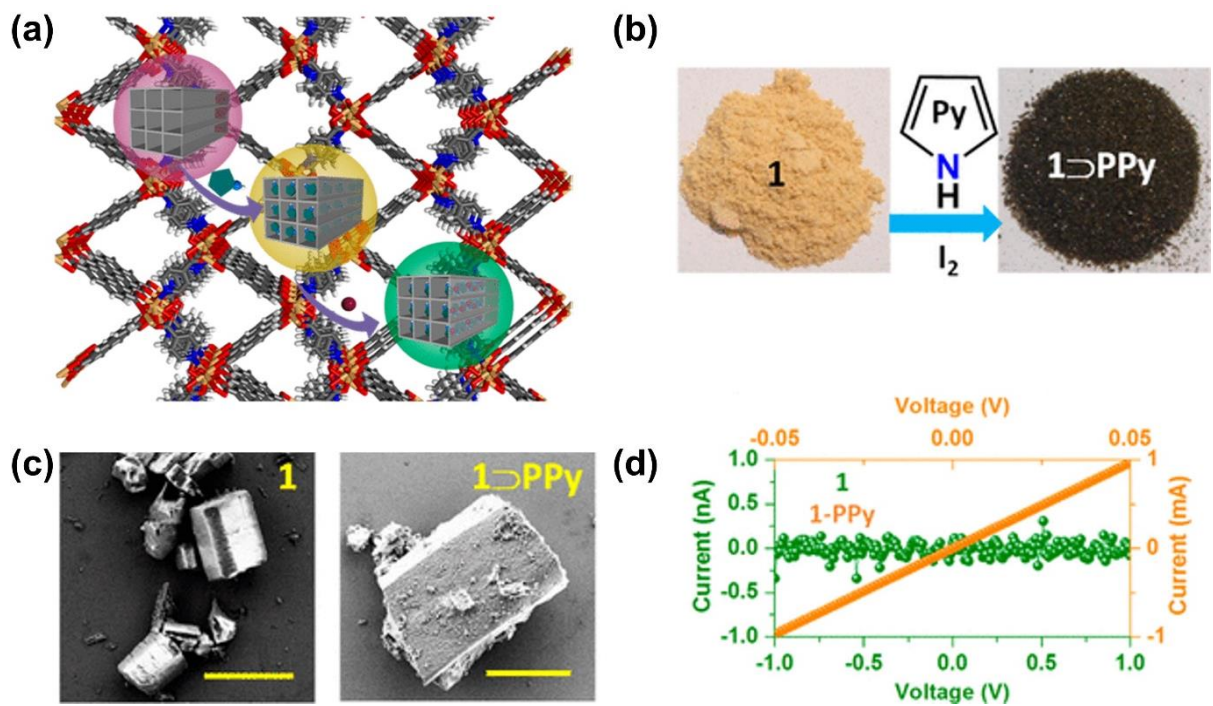

**Figure 9.** (a) Schematic illustration of the inclusion of conductive PPy as guest into the MOFs, with strong host-guest interactions; color code: brown: cadmium, red: oxygen, gray: carbon, white: hydrogen. (b) visual color changes of the pristine MOF upon inclusion of monomer pyrrole and iodine, (c) FE-SEM images of the [Cd(NDC)$_{0.5}$(PCA)] and [Cd(NDC)$_{0.5}$(PCA)]PPy MOF films, (d) comparison of the I-V plots of [Cd(NDC)$_{0.5}$(PCA)] (green circles, pristine MOF) and [Cd(NDC)$_{0.5}$(PCA)]PPy (orange circles, PPy-loaded MOF). Reprinted with permission from Ref.[83] Copyright 2016, American Chemical Society.

The band gap tuning is another tool to boost the electrical conductivity of MOF.[85–88] In this context, we direct readers to the most recently published perspective from the Park group, where the authors comprehensively outline key techniques for investigating the band gap and electronic structure in MOFs and discuss approaches to tune these parameters.[16] A representative example is provided by the study of Ogihara *et al.*, who reported a tunable band gap of 2,6-naphthalene dicarboxylate di-lithium MOF via Li-intercalation.[89] Due to the intercalation, the band gap was reduced to 0.9 eV enabling a hopping mechanism for the charge transport and, consequently, a significant increase in the electrical conductivity **(Fig. 10a)**. Besides, the electrical conductivity of the intercalated MOFs was found to be reversibly dependent on temperature, with the conductive behaviour at 100°C and insulator behaviour at 200°C **(Fig. 10b)**. Thus, this engineered framework can be employed as a prototype switch for protection of electronic devices from overheating. Furthermore, the electrical conductivity of this framework was investigated by electrochemical impedance spectroscopy (EIS) in the frequency range from 100 kHz to 1 Hz. According to the respective Bode plot **(Fig. 10c)**, the resistance of the Li-intercalated MOF was four orders of magnitude lower than that of the pristine framework in the low-frequency regime. The Nyquist plot of the Li-intercalated MOF exhibited a small semi-circle, representing the contributions of both ions and electrons to electrical conductance **(Fig. 10d)**. In contrast, for pristine MOFs, a straight line was found that renders mainly a contribution of ionic conductance. Thus, the EIS studies provide a deep insight into the charge transport phenomena, revealing the contributions of specific charge carriers.[90]

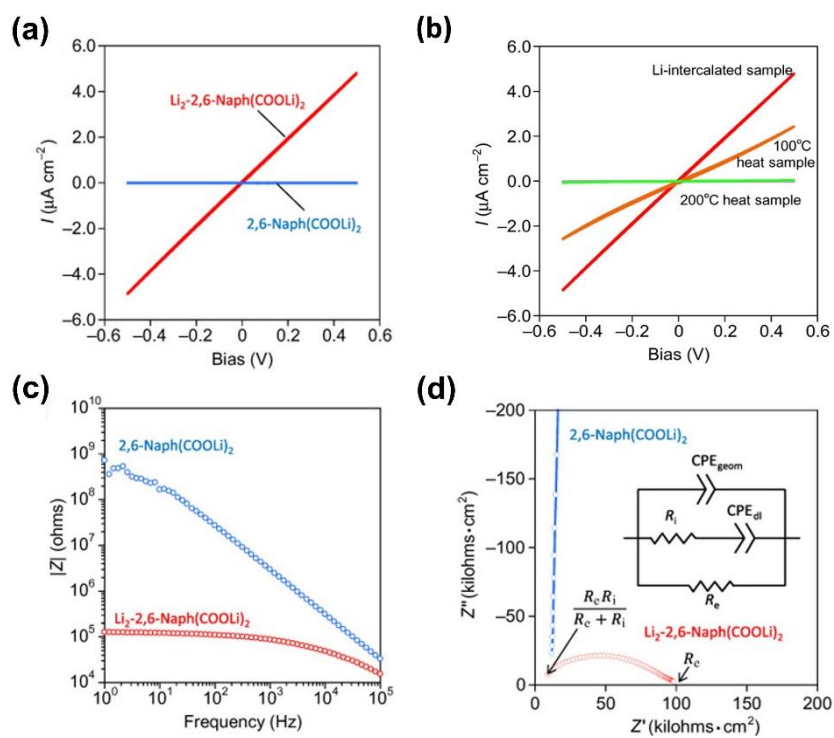

**Figure 10**. (a) Comparison of the *I-V* plots of the pristine (blue line) and Li-intercalated MOFs (red line). (b) Temperature dependent *I-V* measurements. (c) Bode and (d) Nyquist plots for the pristine and Li-intercalated MOFs. The inset in panel (d) represents the corresponding equivalent circuit. Reprinted with permission from Ref.[89] Copyright 2017, the American Association for the Advancement of Science.

Another popular approach to enhance the electrical conductance of MOFs, pioneered by Allendorf and co-authors,[91] is doping of MOFs with redox-active moieties. Along these lines, Liu *et al.* synthesized an array of HKUST-1 SURMOFs with ~45, ~58, and ~71 nm thicknesses by varying the number of cycles in the LbL-LPE method.[34] The films were prepared on Au substrates engineered with a carboxylic-acid-terminated SAM, 9-carboxy-10-(mercaptomethyl) triptycene, denoted as CMMT. Subsequently, these films were loaded with ferrocene (Fc) for conductivity tuning. The pristine and Fc-doped SURMOFs were placed into two-terminal junctions with the gold substrate serving as the bottom electrode and hexadecanethiol (HDT) passivated mercury drop serving as the top electrode **(Fig. 11a)**.

The respective *I-V* curves show a higher current density for the doped SURMOFs compared to the pristine ones as well as an exponential decrease of the current density (and, consequently, an increase in resistance) with the film thickness, corresponding to a very low decay constant ($\beta \approx 0.006$ Å$^{-1}$) (**Fig. 11b-c**). The observed phenomena suggests that the long-range charge transport occurs via an incoherent charge-hopping mechanism. A more substantial effect was observed upon doping of HKUST-1 MOFs with TCNQ and fluorinated TCNQ (F4-TCNQ).[52] The conductivity of HKUST-1 MOF loaded with either TCNQ or F4-TCNQ was found to be at least five orders of magnitude higher than that of the pristine framework. This behavior was further rationalized by an extended hopping transport model which included virtual hops through localized MOF states or molecular super exchange.

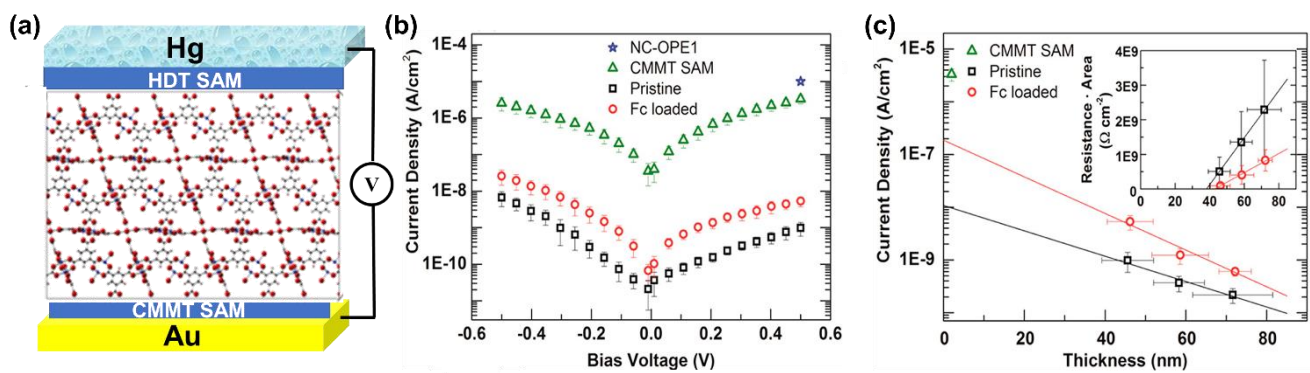

**Figure 11.** (a) Schematic illustration of the two-terminal junction featuring the pristine (shown in the illustration) and Fc-doped HKUST-1 SURMOFs; note that HKUST-1 is formed by copper ion nodes and trimesic acid ligands. Color code: red: oxygen, blue: carbon, and black: copper ion. (b) Representative semi log *J-V* curves for these systems and the SAM-engineered substrate. (c) Semi log plots for dependence of *J* on the SURMOF thickness, showing a linear behaviour; the respective plots for the resistance are presented in the inset. Reprinted with permission from Ref. [34] Copyright 2015, American Chemical Society.

The above experimental results were confirmed and even improved by Jung *et al.* by using a solution shearing method to synthesize a thin film of HKUST-1 MOF with a large area and high crystallinity.[92] This film was photosynthetically doped with TCNQ, resulting in a quantifiable increment of the electrical conductivity by up to seven orders of magnitude. The increase in the electrical conductivity correlated with the soaking time required for the progressive diffusion of TCNQ molecules into the film and their bonding to the metal centers there. Furthermore, the conductivity of the thin films was found to be higher than that of the pellets, which was explained by the nearly defects-free character of these films, favorable for the charge transport. Several research groups have extensively studied the effect of halogen doping on the electrical conductivity of MOFs. Along these lines, Gupta *et al.* investigated the optical, thermoelectric, and semiconducting properties of the Cu[Cu(pdt)$_2$] (pdt =2,3-pyrazinedithiolate) MOFs by infiltration with bromine vapor with varying stoichiometry ratios with respect to Cu ion.[93] Due to such a chemical doping, copper ions in MOFs were partially oxidized, forming a mixed valency state, Br$_x$@Cu$^{II}$[Cu$^{II}$(pdt)$_2$]$_{1-x}$[Cu$^{III}$(pdt)$_2$]$_x$, which facilitates the charge transport via hopping mechanism, resulting in a 10-fold increase in electrical conductivity. Furthermore, for x (Br$_2$ vapor fraction) > 0.5, the pristine MOFs switch from p-type to n-type semiconductor behavior. Since the metal ions are predominantly in the [Cu$^{III}$(pdt)$_2$]$^-$ state at the small x, holes are the dominant charge carriers, corresponding to a p-type semiconductor. In contrast, the doped MOFs feature the [Cu$^{III}$(pdt)$_2$]$^-$ state, which can be easily reduced to [Cu$^{II}$(pdt)$_2$]$^{2-}$, with electrons as majority charge carriers, thus acting as n-type semiconductors.

The electrical conductivity of MOFs can also be modulated by doping with alkali metal ions, as in particular was demonstrated by Zhu and co-workers.[94] In the respective study, pristine MnHHB (HHB = hexahydroxybenzene) and MnTHBQ (THBQ = tetrahydroxy-1,4-benzoquinone) MOFs were doped with metal ions (Rb$^+$ or Cs$^+$) resulting in the MnRbHHB/THBQ and MnCsHHB/THBQ frameworks, respectively. For the electrical conductivity measurements, the MOFs, produced initially as powders, were pressed to make pellets which were then assessed electrically by four-probe method. In all cases, an increase in the electrical conductivity upon increasing temperature (from 300K to 400K) was observed, which indicated a semiconducting nature of the MOFs (**Fig. 12a**). For instance, the electrical conductivity (σ) of the MnHHB MOFs was found to be $5.4 \times 10^{-3}$ S cm$^{-1}$ at 300 K and $1.8 \times 10^{-2}$ S cm$^{-1}$ at 400 K. In contrast, the MnRbTHBQ and MnCsTHBQ MOFs exhibited lower conductivity of $1.6 \times 10^{-7}$ S cm$^{-1}$ and $1.2 \times 10^{-8}$ S cm$^{-1}$ at 300 K and $7.5 \times 10^{-6}$ S cm$^{-1}$ and $4.8 \times 10^{-7}$ S cm$^{-1}$ at 400 K. Using the Arrhenius equation, $\sigma = \sigma_0 \exp(-E_a/k_B T)$, with $\sigma_0$ as prefactor, the activation energy for the charge transport ($E_a$) was determined at 0.128, 0.39 and 0.34 eV for MnHHB, MnRbTHBQ, and MnCsTHBQ, respectively **(Fig. 12b)**. The π-π stacking distance for the MnRbTHBQ and MnCsTHBQ was estimated at ~3Å and for the MnHHB at ~2.93Å. It is significant that only a ~5% difference in the stacking distance results in a 100-fold enhancement in the electrical conductivity for MnHHB compared to MnRbTHBQ and MnCsTHBQ, relying on the stronger π-π interaction in the former framework.

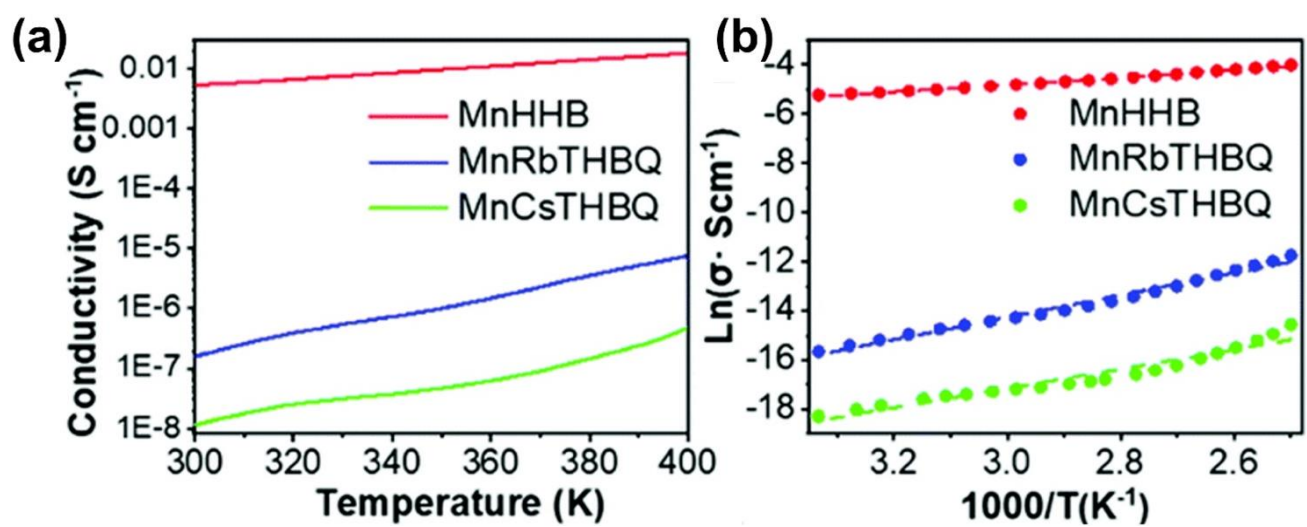

**Figure 12.** (a) Plots of the electrical conductivity as a function of temperature for three different MOFs, (b) Arrhenius-like plots of lnσ as a function of 1000/T. Reprinted with permission from Ref.[94] Copyright 2022, The Royal Society of Chemistry.

# 8. Solvent-induced conductivity modulation

A specific kind of doping are polar solvent molecules that can modulate many properties of materials including electrical conductivity.[95,96] For instance, coordination of *N, N*-dimethylformamide (DMF) solvent to the unsaturated Fe centers of the $Fe_2$(DSBDC) (DSBDC = 2,5-disulfidobenzene-1,4-dicarboxylate) MOF enhances the electrical conductivity of this framework by a factor of ~$10^3$, from $10^{-9}$ S cm$^{-1}$ to $10^{-6}$ Scm$^{-1}$.[97] Note that the electrical conductivity was measured using the paste-wire method, in which the pellet was sandwiched between the carbon paste and gold wire. The increase in the conductivity was attributed to the fractional electron transfer from the Fe center to the polar solvent molecules. Theoretical calculations revealed that the ionization potential of DMF is lower than the work function of the $Fe_2$(DSBDC) MOF by nearly 1 eV, and accordingly, electron transfer takes place when DMF is introduced into the MOFs.

# 9. Metal-catecholate based MOFs

The metal-catecholates (M-CAT) based MOFs exhibit strong charge delocalization due to the significant overlap of electronic states between the metal nodes and organic linkers. The combination of in-plane conductivity and layered structure makes them an excellent platform for creating ultrathin MOF films, suitable both for model studies of electrical conductivity and device fabrication. However, the conductivity of these films turned out to be much lower (by up to three orders of magnitude) than that of the respective single-crystal MOFs, making it crucial to increase their crystallinity and orientational order. In this context, Medina and co-workers prepared well-oriented M-CAT-1 thin films with thicknesses of 180-200 nm by using vapor-assisted conversion technique. These films, comprised of 2,3,6,7,10,11-hexahydroxytriphenylene (HHTP) linkers and $Ni^{2+}$, $Co^{2+}$, and $Cu^{2+}$ nodes, were subjected to electrical measurements and integrated into model solar cells between ITO and Al electrodes.[98] Both $Ni^{2+}$ and $Co^{2+}$ based MOFs showed symmetric current-voltage responses in the voltage ranges of ± 5 V, as measured by van der Pauw methods with a contact distance of 5 mm. The corresponding electrical conductivity values were $1.1 \times 10^{-3}$ S cm$^{-1}$ and $3.3 \times 10^{-3}$ S cm$^{-1}$, respectively. The respective solar cells showed diode-type behavior under AM 1.5G light illumination. It was assumed that the well-ordered stack of molecular layers in the oriented thin films serves as an effective conduction path for the photogenerated charge carriers.

An alternative approach to the synthesis of highly oriented, ultrathin (10 nm) Cu-CAT-1 MOF films (sequential transfer of nanosheets) was reported by Carlos and co-workers, resulting in a relatively high electrical conductance.[99] For the charge transport studies, a thin film-based device with bottom-contact geometry was fabricated, using pre-patterned Au electrodes as the source and drain electrodes and $SiO_2$ substrate and highly doped n-type Si as dielectric and gate electrodes, respectively (**Fig. 13a**). Optical microscopy confirms the full device coverage after the transfer of the Cu-CAT films onto the electrode template (**Fig. 13b**). A linear increment in the conductivity as a function of applied bias at room temperature reveals an ohmic behaviour of the MOFs (**Fig. 13c**). The film conductivity at room temperature ($\sigma$) was calculated to be ~$10^{-4}$ S cm$^{-1}$ which is much lower than that of the respective single crystals ($\sigma = 2 \times 10^{-1}$ S cm$^{-1}$). The activation energy, obtained from the temperature-dependent electrical conductivity data (**Fig. 13d**), was estimated at 0.24 eV. The devices showed a nonlinear relationship between $\ln(\sigma(T)/\sigma(RT))$ and T which is common for conducting polymers and could be rationalized by the coexistence of two different conduction mechanisms. Specifically, at relatively high temperatures (T > 240 K), the electrical conduction is dominated by thermally activated charge carriers in view of the relatively narrow fundamental band gap (~0.48 eV), which is double of the activation energy ($E_g = 2E_a$). In contrast, at lower temperatures (T < 240 K), the conduction is governed by a hopping mechanism, which was further confirmed by the observed $T^{-1/4}$ relationship, typical of hopping in Mott's 3D systems. It was assumed that the Cu-CAT-1 MOFs feature both inter- and intralayer charge transport relying on the π–π interactions within the framework. Note that the same authors have recently linked carbon nanotubes (CNT) to the ligands of UiO-68-TZDC MOF (TZDC = 4,4'-(1,2,4,5-tetrazine-3,6-diyl)dibenzoic acid) resulting in a room temperature conductivity of the hybrid MOF-CNT wrap of $4 \times 10^{-2}$ Scm$^{-1}$.[100]

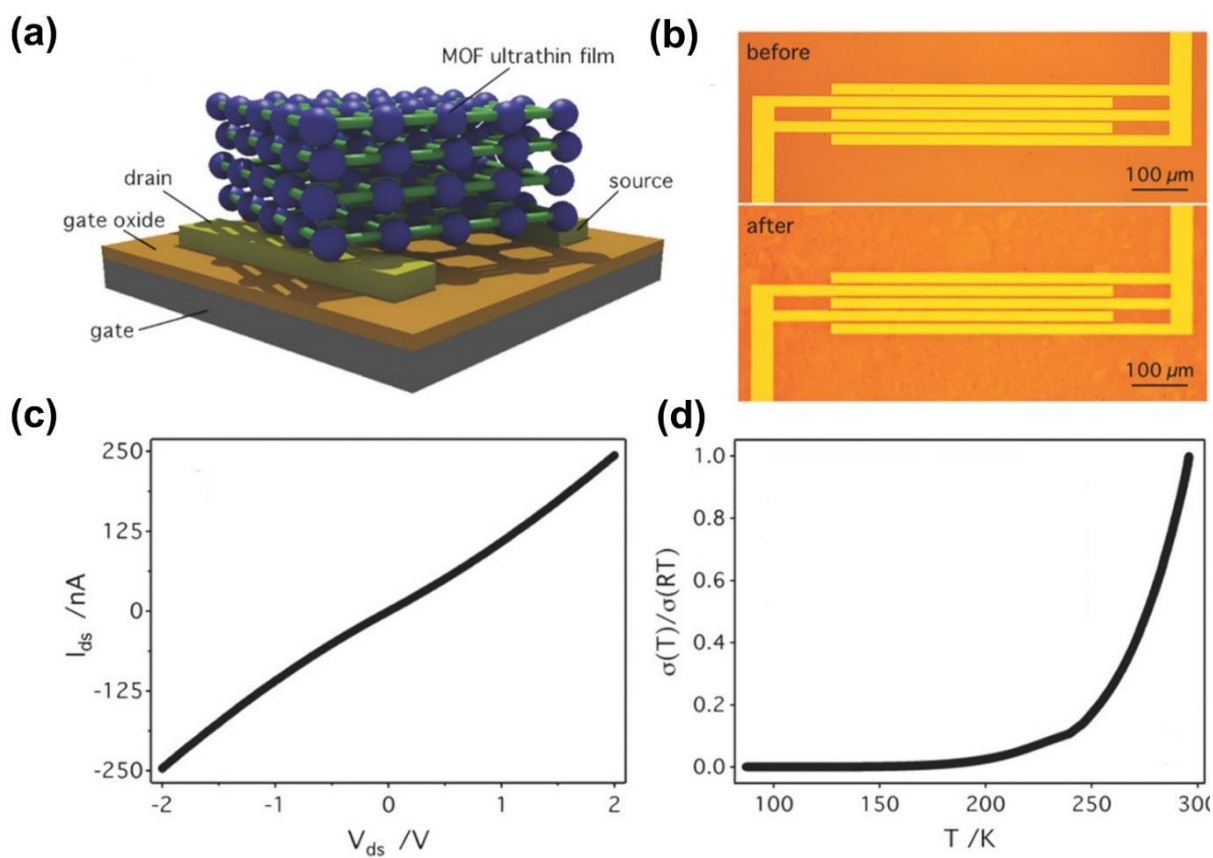

**Figure 13.** (a) Schematic illustration of the Cu-CAT-based device with bottom-contact geometry; the nodes (copper ions) and linkers (catecholates) are drawn in blue and green, respectively. (b) Optical microscopy images

of the real device before and after the transfer of the Cu-CAT-1 films onto the electrodes. (c) representative *I-V* curve of the device, measured at room temperature. (d) Dependence of the relative electrical conductance on temperature. Reprinted with permission from Ref.[99] Copyright 2018, Wiley-VCH.

Recently, Dincă and coworkers reported tuning of the band gap of metal-alloyed conjugated MOFs $(M_xM'_{3-x})(HITP)_2$ (MM' = CuNi, CoNi, CoCu; HITP = 2,3,6,7,10,11-hexaiminotriphenylene) by varying the metal ions ratio, leading to a considerable increase in electrical conductivity.[62] Later, Dong and coworkers synthesized a novel wavy 2D c-MOF [$Cu_3(HFcHBC)_2$] composed of $Cu^{2+}$ and nonplanar ligand 2,3,10,11,18,19-hexafluoro-6,7,14,15,22,23-hexahydroxycHBC (HFcHBC) followed by electrical conductivity measurement.[101]

## 9.1 Ni-HITP MOFs

Promising systems are 2D conductive MOFs, such as $Ni_3(HITP)_2$, which consist of 2,3,6,7,10,11-hexaaminotriphenylene (HITP) linkers and nickel ions and show high bulk electrical conductivity (> 5000 Sm$^{-1}$). These frameworks can be particularly useful in the context of supercapacitor and battery applications and electrocatalysis.[102–104] The electrical conductance of these MOFs was in particular studied in the Si/SiO$_2$/MOFs/Ti/Pd device configuration.[3] The rod-shaped nanocrystals of $Ni_3(HITP)_2$ with a length of ~2 μm and a diameter of ~200 nm were drop-casted onto Si/SiO$_2$ substrate, followed by e-beam evaporation of Ti/Pd top contacts. The resulting devices showed electrical conductance (G) of 1.3 mS at 295 K and 0.7 mS at 1.4 K when measured by four-probe method and 0.25 mS at 295 K and 0.13 mS at 1.4 K when measured by two-probe method. The electrical conductance of the single crystal MOFs was found higher than that of the respective polycrystalline thin films, which is mostly related to the grain boundary contribution. Later, Ohata *et al.* synthesized well-oriented thin films of $Ni_3(HITP)$ MOFs by employing air/liquid bottom-up approach.[105] This work is one of the first reports on synthesizing crystalline thin MOF films with a unidirectional orientation, as ensured by synchrotron-based X-ray crystallography. These thin films can be easily transferred onto any substrate which makes them superior to the frameworks fabricated by solvothermal or vapor-induced conversion method in which MOFs can be grown on a specific substrate only. The in-plane electrical conductivity was measured using a thin film device with two Au electrodes and found to be 0.6 S cm$^{-1}$ for the 70 nm films, which was the highest of the previously reported values for nanosheets of similar thickness. The respective *I-V* plot had a linear character, suggesting the Ohmic behaviour of the electrical conductivity in the $Ni_3(HITP)$ films, which was ascribed to the high orientation order in these systems. The above results pave the way for the fabrication of highly efficient electronic devices featuring highly oriented MOF nanosheets. Along with the Ni-HITP MOFs, electrically conductive MOF constructed with nickel and 2,3,6,7,10,11-hexahydroxytriphenylene (HHTP) were recently reported.[106] These frameworks, grown by electrosynthesis, feature disc- and flower-like morphologies. The electrochemical performance of the flower-shaped frameworks greatly (up to 10 times) exceeds that of bulk Ni-HHTP, both in terms of gravimetric capacitances and electrochemically active surface areas.

## 9.2 Ga-based MOFs

To date, the utmost efforts in modulating the electrical conductivity of MOFs have been centred around the tuning of ligand structures. In contrast, the role of metal ions in influencing conductivity has gained comparatively less attention. The electrically conducting MOFs that have been reported consist mainly of only a few metal ions, e.g., Ni, Cu, and Co. Notably, Dincă and co-workers have expanded the library of electrically conducting MOFs through the pioneering synthesis of Ga metal ions-based frameworks.[107] They employed Ga (iii) as a metal node and 2,3,6,7,10,11-hexahydroxytriphenylene as a ligand. The conductivity of the synthesized MOF was found to be more than 3 mS/cm, which is well comparable to the earlier reported homologous frameworks based on the Ni(II) and Co(II) metal ions. The authors inferred that charge transport within the Ga-based MOFs is predominantly facilitated through π–π stacking interactions, leading to high electrical conductivity, similar to the Ni(II) and Co(II) case. This finding marks a significant breakthrough and opens new avenues for utilizing π-block elements in MOF synthesis.

## 10. Lanthanide MOFs

Another promising alternative to the "standard" MOF systems exploiting predominantly Ni, Cu, and Co nodes is provided by lanthanide MOFs. To this end, Dincă and co-workers studied the conductivity relationship in a series of Ln-HHTP MOFs comprising HHTP ligands and lanthanide cations (Ln = Yb, Ho, Nd, and La), showing that the emergence of high conductivity in layered MOFs does not necessarily require a metal-ligand bond with highly covalent character and that interactions between organic ligands alone can produce efficient charge transport pathways.[108] The crystallographic structure of these MOFs is illustrated in **Fig. 14a-c**. As to the electrical properties, Yb-HHTP and Ho-HHTP, which feature comparatively small metal ions and have more compact stacking, exhibited, as shown in **Fig. 14d**, much higher electrical conductivity than that of La-HHTP and Nd-HHTP. The structural data reveal that the conductivity was dominated by the stacking distance rather than by the extent of covalency between the respective ions and ligands. This result suggests that the charge transport takes place out-of-the plane instead of in-plane. Note that all Ln-HHTP MOFs showed a similar activation energy of 250 meV in the temperature range 225–300 K, calculated from the Arrhenius equation.

In view of the dominant role of the stacking distance in the lanthanide MOFs, tetrathiafulvalene tetrabenzoate (TTFTB), known for strong π–π interactions, was considered as a linker. To this end, Dincă and co-workers synthesized three different MOFs composed of $La^{3+}$-TTFTB and denoted as **1**, **2**, and **3** via the conventional solvothermal method, varying temperature and composition of the solvent ($H_2O$: DMF: EtOH) (**Fig. 15a**).[109] The strength of the π–π interactions in these MOFs, dictating the electrical conductivity, was mainly defined by S…S contact distance between the adjacent ligands, which varied across the series in the order of **1** < **2** < **3** and was estimated at 3.60, 4.08, 7.07 Å for **1**, **2**, and **3**, respectively. The electrical measurements were performed by sandwiching the $La^{3+}$-TTFTB MOF pellets between two electrical contacts, viz. carbon paste (bottom contact) and gold wire (top contact), as schematically shown in **Fig. 15b**. The charge transfer rate was found to be directly proportional to the extent of π-π overlapping in TTFTB which resulted in the conductivity order of $2.5(7) \times 10^{-6}$ S cm$^{-1}$, $9(4) \times 10^{-7}$ S cm$^{-1}$, and $1.0(5) \times 10^{-9}$ S cm$^{-1}$ for **1**, **2**, and **3**, respectively. It was found that the charge transport in these MOFs was mediated by thermally activated hopping mechanism with activation energies of 0.28, 0.20, and 0.44 eV for **1**, **2**, and **3** respectively.

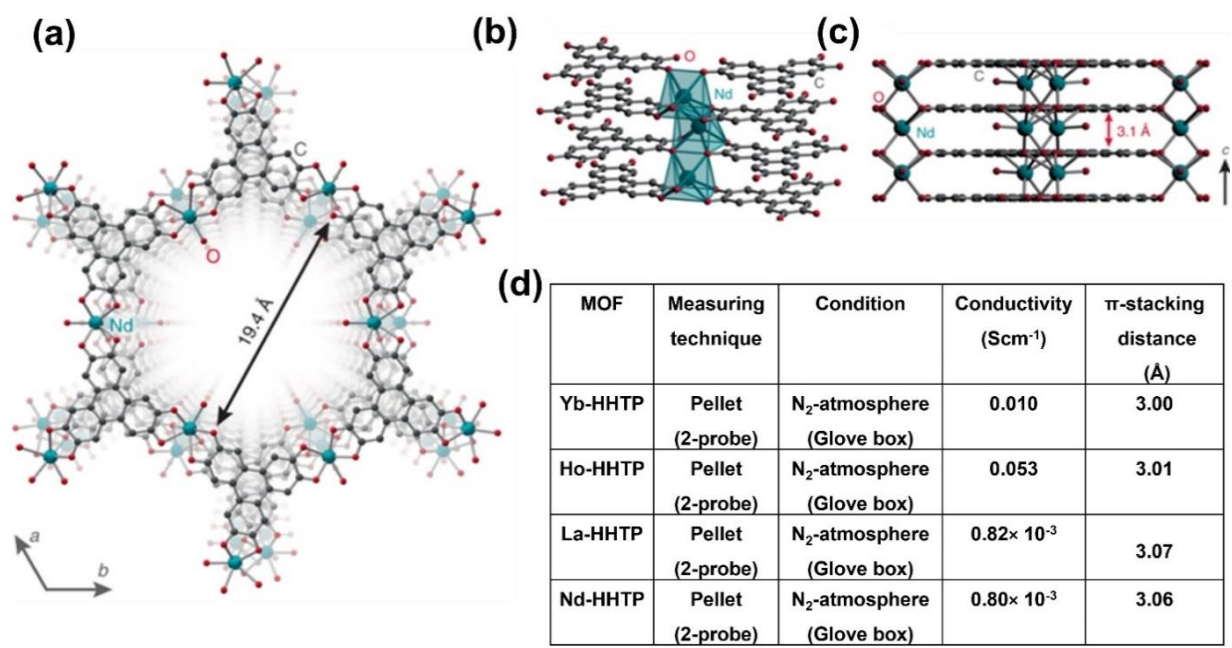

**Figure 14.** (a) Crystallographic structure of Nd-HHTP MOF (top view) with the marked pore diameter; (b) Side view of the same structure emphasizing Nd atoms connected with the ligands. (c) Respective interlayers with a stacking distance of 3.1 Å. (d) Electrical conductivity values of the Ln-HHTP MOFs (Ln =Yb, Ho, Nd, and La). The color codes for the panels (a-c) are given in the respective panels. Reprinted with permission from Ref.[108] Copyright 2020, Springer Nature.

Complementary to the above frameworks, Skorupskii and Dincă recently synthesized intrinsically electrically conductive and structurally isotropic $Eu_6HOTP_2$ and $Y_6HOTP_2$ MOFs, consisting of 2,3,6,7,10,11-hexaoxytriphenylene (HOTP) linkers and Eu and Y metal nodes.[110] These MOFs showed electrical conductivity on the order of $10^{-6}$ to $10^{-5}$ S/cm.

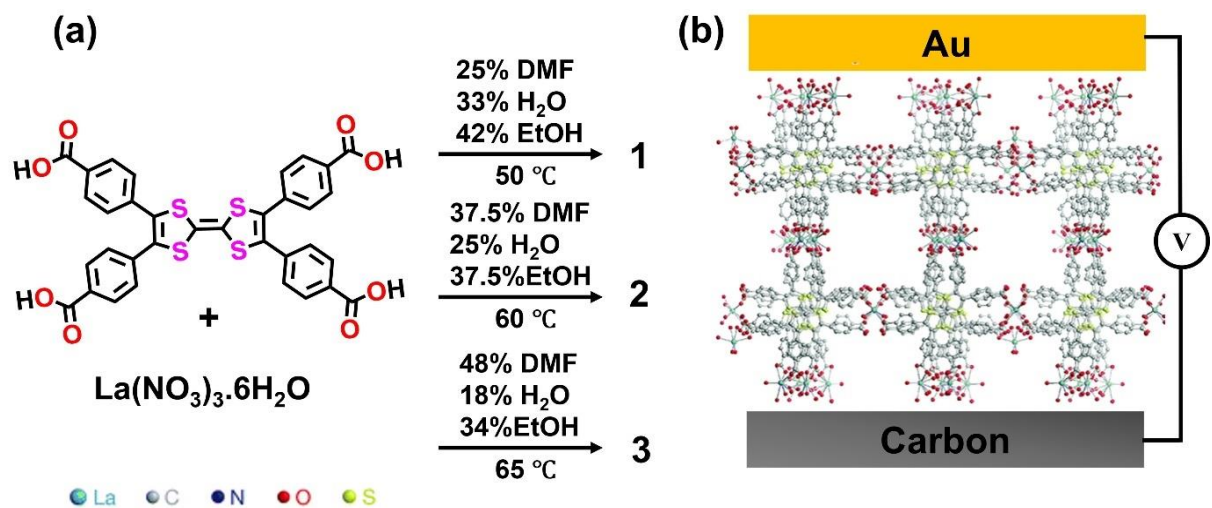

**Figure 15.** (a) Schematic of the synthesis of different La$^3$-TTFTB MOFs (**1, 2,** and **3**) by variation of temperature and solvent composition. (b) Schematic illustration of MOFs-based device with an Au/H$_4$TTFB/carbon configuration; color code: light blue: lanthanide, gray: carbon, dark blue: nitrogen, red: oxygen, yellow: sulfur. Reprinted with permission from Ref.[109] Copyright 2019, The Royal Society of Chemistry.

## 11. MOF-based devices for specific electronic applications

Research on MOFs-based electronics has drastically advanced in recent years, targeting photoelectric and memory devices, field-effect transistors (FETs), rectifiers, etc.[111,112] For instance, Yoon *et al.* reported the first example of MOF-based resistive memory devices using γ-cyclodextrin based Rb-CD-MOFs.[113] Recently, Pham and co-workers studied volatile threshold as well as non-volatile memory devices using zirconium(IV)-carboxylate-based MOFs (UiO-66). These devices showed excellent characteristics, such as a reasonable ON/OFF current ratio (~10$^4$), high endurance (5 × 10$^2$ cycles), and long-time memory retention (10$^4$ s).[114] Complementary, Ballav and co-workers designed and fabricated a rectifier on the MOF basis.[115] Within the respective project, these authors prepared a thin film (a thickness of ~700 nm) of Cu(II)-1,2,4,5-benzene tetracarboxylic acid (Cu-BTEC) on the gold substrate by LbL method (**Fig. 16a-b**) and doped subsequently a part of this film with TCNQ, creating an electronic heterostructure resembling a typical p-n junction. The rectification ratio (RR) for the resulting device exceeded 10$^5$ (**Fig. 16c**), which is one of the best values reported so far for molecular and hybrid rectifiers.[45] In addition, the same group recently reported heterostructure MOFs composed of Cu$_3$BTC$_2$ (top-layer) and TCNQ@Cu$_3$BTC$_2$ (bottom layer) grown on pre-functionalized Au-coated PET substrate.[116] These frameworks showed a RR ≈ 10$^5$ and featured a high charge carrier concentration of ca. 770 × 10$^{15}$ cm$^{-2}$ and a high charge carrier mobility of 22.7 cm$^2$ V$^{-1}$ S$^{-1}$.

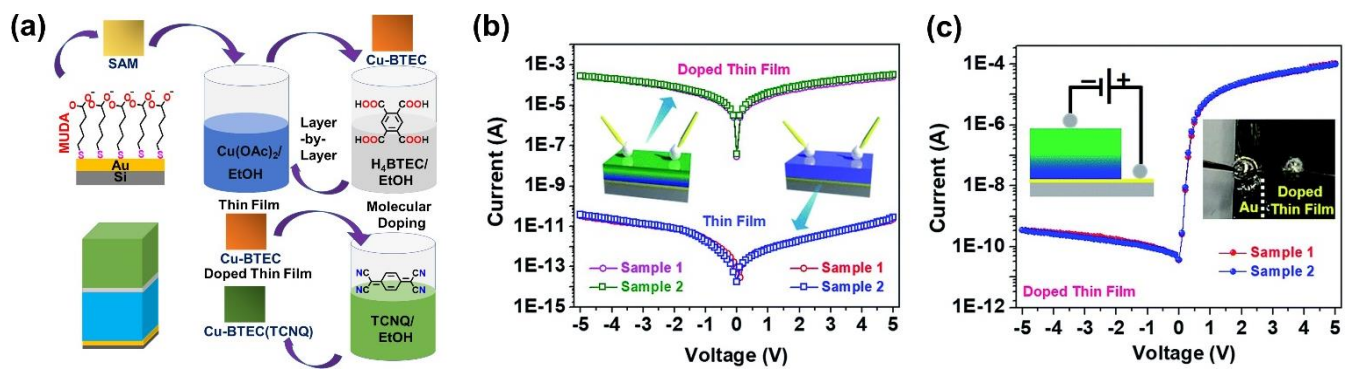

**Figure 16.** (a) Schematic view of the fabrication steps of the Cu–BTEC thin film by LPE-LbL method (ten repeating cycles) and its subsequent doping with TCNQ. (b) *I-V* plots (in-plane measurements) for undoped (blue and red) and doped (green and magenta) Cu-BTEC thin film. A considerable enhancement in the electrical conductivity was obtained upon doping. (c) *I-V* plots (blue and red) for the resulting device (cross-plane measurement) showing an RR value of ≥$10^5$. Reprinted with permission from Ref.[115] Copyright 2019, The Royal Society of Chemistry.

## 12. Conclusions and future perspectives

This review discusses recent advances made with diverse bulk MOFs and MOF thin films in context of controlling and tuning their electric properties, putting particular attention on the relevant fabrication strategies, anisotropy of conductance, most promising types of conductive MOFs, and prototype electronic devices using selective MOFs. However, despite a variety of promising results, integration of MOFs into real electronic devices is still a challenge, along with many other challenges and unresolved questions in the MOFtronics that need further studies and better understanding. A comprehensive list of the respective challenges and questions, most of which are still relevant, can be found in the recent review by Dincă and co-authors.[12] This list includes disentangling the properties of charge carriers in MOFs, realizing more examples of isotropic conduction pathways in MOFs, further development of mixed-metal and mixed-linker systems, further elucidation of transport mechanisms relevant for 2D MOFs, isolation and characterization of single layers of 2D MOFs, expanding the library of ligands for MOFs featuring through-space charge transport pathways, more detailed structural investigations of host-guest interactions in MOFs, influence of environment and atmosphere on MOF conductivity values, and a better documentation of the materials preparation and measurement routes.[12] In our opinion and in the context of the present review, this list can be complemented by the following points:

i. *Fabrication of heterostructure MOFs:* Heterostructure MOFs, comprised of either two different MOFs or differently doped MOFs (such as in Ref.[116]) would be highly important for applications. They can in particular behave as electrical diodes, relying on the different electrical conductance properties of the contributing parts within the entire device configuration, such as electrode/MOF1/MOF2/electrode or electrode/doped-MOF1/MOF1/electrode. Also, reversible switching of rectifying to nonrectifying current across specifically designed heterostructure MOFs becomes possible.[117]

ii. ***Variation of MOF thickness***: To our surprise, charge-transport behavior for different thicknesses of the same MOFs have been explored so far to a very limited extent only (see, e.g., Ref[34] and Ref[45]). However, such studies are necessary to derive additional knowledge regarding the relevant charge transport mechanism and to optimize the thickness of the MOF films for specific applications.

iii. ***Role of electrodes in device fabrication***: It is indisputable that electrodes play an important role in controlling charge transport properties of any electronic device, including MOF-based ones. In particular, different electrodes have most likely different coupling efficiencies with the MOF and different Fermi levels ($E_F$), affecting the heights of the injection barriers at the respective interfaces ($E_{HOMO/LUMO}-E_F$) and, consequently, controlling the charge injection at these interfaces.[118] Generally, reference measurements with identical but variable electrodes as bottom and top contacts would be useful to distinguish between their impact and the contribution of MOF itself.

iv. ***Metal-free electrodes in device stacking***: Metal-free electrodes are highly desirable for real-world applications as metals are costly and usually require a high vacuum deposition system. In this context, conductive carbon or graphene can be an excellent alternative to replace metal electrodes.[119–122] In particular, conductive carbon has a Fermi level like Au but is cost-effective, patternable, and suitable for flexible substrates. It can also impede filament formation occurring frequently in the course of top metal contact deposition.

v. ***Impedance spectroscopy studies on MOFs devices***: Further efforts are required to study and rationalize specific electrical parameters of MOF-based electrical devices, such as contact resistance, charge-transfer resistance, and capacitance. These parameters are important from both fundamental viewpoint and for MOF-based electronics since their knowledge helps to understand and distinguish their specific roles in the conduction mechanism. Conventional DC-based measurements do not deliver these parameters in contrast to impedance spectroscopy providing them in relation to a suitable equivalent circuit.

vi. ***In-depth electrochemical studies on MOFs***: Adequate electrochemical measurements are necessary to study the redox-conductivity of MOFs. The reliable electrochemical analysis would facilitate understanding their double-layer capacitance, charge-transfer rate constant, etc. To date, just several redox-active MOFs are known[11,123–125] but recent theoretical and experimental work suggests that these MOFs can be good electrode materials.[126]

vii. ***Statistics and device stability:*** Regretfully, many reports do not pay sufficient attention to statistics of the charge transport data (number of working junctions out of total measured junctions, etc.) and even less studies provide information about the stability of MOF-based electronic devices. This information is however important to estimate the reliability of the presented data and the real value and usability of the devices.

viii. *Charge transport mechanism:* It is crucial to understand in-depth the charge transport in MOF-based electronic devices including such relevant parameters like barrier height, electrode-MOF coupling, and intermolecular coupling. For instance, the charge injection barrier height, defined by $E_{HOMO/LUMO}$-$E_F$, can be minimized by choosing an appropriate electrode (known $E_F$) for the known values of HOMO and LUMO. Also, the standard *I-V* representation of the charge transport data can be well complemented by Ln$J$ vs $V$, ln$J$ vs $E$, or Ln$J$ vs $E^{1/2}$ ($J$, and $E$ stand for the current density and electric field, respectively) plots, which can help to decide whether the charge transport is electric filed-dependent or not.

ix. *Theoretical models*: Simulations and suitable theoretical models can help to understand better the physical and chemical factors governing the electrical conductivity of MOFs. In addition to these fundamental issues, theoretical simulations can also open new ways for the rational design of MOFs, beyond time- and resource-consuming trial-and-error experimental approach, which is frequently applied for optimization of new systems.

x. *Adapting of design ideas*: Novel design ideas can be implemented. One example, is the introduction of built-in electric fields into MOFs, on the basis of dipolar linkers,[127] following the strategy of embedded dipole adapted from molecular self-assembly.[128,129] This approach creates possibilities for band structure engineering in MOFs and may also be used to tune the transfer of charges from donors loaded via programmed assembly into MOF pores.[127] Another option is the introduction of chiral linkers providing potentially spin-dependent charge transport, following the concept of chirality-induced spin selectivity (CISS).[130–133] Alternative possibility of achiving spin selectivity in MOFs is the use of magnetic molecular building blocks, following ideas from molecular electronics.[134]

Despite many examples of exciting research work, we think that significant efforts in design, characterization, and understanding of functional MOFs and SURMOFs are still needed to establish and commercialize them as building blocks of real-world electronic devices. Hopefully, this review will motivate the MOF community to further efforts and systematic studies of the charge transport phenomena in diverse MOF systems, both in the context of fundamental science and applications.

## List of Abbreviations

| | |
|---|---|
| A | Junctions area (in cm$^2$) |
| AC | Alternating current |
| ALD | Atomic layer deposition |
| BDC | 1,4-benzene dicarboxylate |
| CAFM | Conducting atomic force microscopy |
| CB | Conduction band |
| Cdl | Double layer capacitance (Farad) |
| CFSE | Crystal Field Stabilization Energy |
| CMOS | Complementary metal-oxide–semiconductor |
| CP | Conjugated polymers |
| CPs | Coordination polymers |
| CV | Cyclic voltammetry |
| CVD | Chemical vapor deposition |
| DC | Direct current |
| DFT | Density function theory |
| DMF | Dimethyl formamide |
| DOS | Density of states |
| DSBDC | 2,5-disulfidobenzene-1,4-dicarboxylate |
| d | Thickness of the molecular layers between two conductors |
| ECD | Elctrochromic devices |
| ECMOFs | Electrically conductive metal-organic frameworks |
| E-field | Electric field (V/cm) |
| E/M/E | Electrode/molecules/electrode |
| EIS | Electrical impedance spectroscopy |
| ET | Electron transfer |
| $E_a$ | Activation energy (eV) |
| $E_F$ | Fermi levels (eV) |
| $E_g$ | Band gap energy (eV) |
| Fc | Ferrocene [Fe(cyclopentadiene)$_2$] |
| FMOs | Frontier molecular orbitals |
| FTO | Fluorine doped tin oxide |
| FESEM | Field-emission electron microscopy |
| HHTP | 2,3,6,7,10,11-hexahydroxytriphenylene |
| HKUST | Hong Kong University of Science and Technology |
| HOMO | Highest occupied molecular orbitals |
| iMOFs | Intercalated metal-organic frameworks |
| ITO | Indium tin oxide |
| J | Current density (Amp/cm$^2$) |
| $J_o$ | pre-factor |
| $K_B$ | Boltzmann constant |
| $k_{et}$ | Electron transfer rate |
| LbL | Layer-by-layer |
| LPE | Liquid-phase epitaxy |
| LUMO | Lowest unoccupied molecular orbitals |
| M-CAT-1 | metal-catecholate |
| M/I/M | Metal/insulator/metal |
| MJs | Molecular junctions |
| MOFs | Metal-organic frameworks |
| PET | Polyethylene terephthalate |
| PPF | Pyrolyzed photoresist film |
| PVD | Physical vapor deposition |
| PEDOT | Poly(3,4-ethylenedioxythiophene) |
| R | Resistance (in Ohm) |
| RCA | Radio Corporation of America |
| $R_{ct}$ | Charge-transfer (or molecular) resistance (in Ohm) |
| $R_s$ | Contact resistance (in Ohm) |
| S | Siemence (Ohm$^{-1}$) |
| SAMs | Self-assembled monolayers |
| SCPs | Surface coordination polymers |
| SEM | Scanning electron microscopy |
| SURMOFs | Surface-confined metal–organic frameworks |
| T | Temperature (in K) |
| TCNQ | 7,7,8,8-Tetracyanoquinodimethane |
| TDDFT | Time-dependent density functional theory |
| THT | 2,3,6,7,10,11-tripheylenehexathiolate |
| TTFTB | tetrathiafulvalene tetrabenzoate |
| UPS | Ultraviolet photoelectron spectroscopy |
| VAC | Vapor-assisted conversion |
| VB | Valence band |
| vdP | Van der Pauw |
| WF | Work function (eV) |
| XPS | X-ray photoelectron spectroscopy |
| ZIF | Zeolitic imidazolate framework |
| $\alpha$ | Potential well height (eV) |
| $\beta$ | Charge-transport exponential decay constant |
| $\sigma$ | Electrical conductivity (S cm$^{-1}$) |
| $\phi_T$ | Energy barrier for coherent tunneling (eV) |

**Declaration of competing interest:**

The authors declare no competing interest in this review article.

**Acknowledgments**

RKP gratefully acknowledge the junior research fellowships from the Council of Scientific and Industrial Research, New Delhi. PJ acknowledges IIT Kanpur for providing Institute postdoctoral fellowship. PCM acknowledges financial support from the Council of Scientific & Industrial Research, project NO.:01(3049)/21/EMR-II, New Delhi, India and the Science and Engineering Research Board (Grant No. CRG/2022/005325), New Delhi, India. The authors thank Dr. Vikram Singh for fruitful discussions.